\documentstyle[12pt]{article}
\begin{document}

\newcommand{\nc}{\newcommand}

\nc{\eqn}[1]{Eq.~\ref{eq:#1}}
\nc{\eqns}[2]{Eqs.~\ref{eq:#1} and \ref{eq:#2}}
\nc{\fig}[1]{Fig.~\ref{fig:#1}}
\nc{\figs}[2]{Figs.~\ref{fig:#1} and \ref{fig:#2}}
\input epsf
\nc{\figdir}{} 
\nc{\figmac}[5]{\begin{figure}
\centerline{\parbox[t]{#1in}{\epsfbox{\figdir #2.ps}}}
\caption[#4]{\label{fig:#3} #5}\end{figure}}
%
\def\epsfsize#1#2{\ifdim#1>\hsize\hsize\else#1\fi}

\renewcommand{\slash}[1]{/\kern-7pt #1}
\newcommand{\beq}{\begin{equation}}
\newcommand{\eeq}{\end{equation}}
\newcommand{\beqn}{\begin{eqnarray}}
\newcommand{\eeqn}{\end{eqnarray}}
\newcommand{\beqns}{\begin{eqnarray*}}
\newcommand{\eeqns}{\end{eqnarray*}}
\newcommand{\nn}{\nonumber}

\def\cA{{\cal A}}
\def\cB{{\cal B}}
\def\cC{{\cal C}}
\def\cD{{\cal D}}
\def\cE{{\cal E}}
\def\cF{{\cal F}}
\def\cG{{\cal G}}
\def\cH{{\cal H}}
\def\cI{{\cal I}}
\def\cK{{\cal K}}
\def\cM{{\cal M}}
\def\cN{{\cal N}}
\def\cO{{\cal O}}
\def\cP{{\cal P}}
\def\cR{{\cal R}}
\def\cS{{\cal S}}
\def\cX{{\cal X}}
\def\cZ{{\cal Z}}

\def\qb{\bar{q}}
\def\kb{\bar{k}}
\def\rb{\bar{r}}
\def\Qb{\bar{Q}}
\def\Kb{\bar{K}}
\def\Rb{\bar{R}}

\def\to{\rightarrow}
\newcommand{\lra}{\leftrightarrow}
\newcommand{\la}{\langle}
\newcommand{\ra}{\rangle}
\def\A#1#2{\la#1#2\ra}
\def\B#1#2{[#1#2]}
\def\s#1#2{s_{#1#2}}
\def\h#1#2#3#4{\la#1#2#3#4\ra}
\def\P#1#2{{\cal P}_{#1#2}}
\def\L#1#2{\left\{#2\right\}_{#1}}
\newcommand{\Lmin}{{\rm L}}
\def\Li{{\rm Li_2}}
\def\ms{$\overline{{\rm MS}}$}

\newcommand{\y}{\gamma}
\newcommand{\yf}{\gamma_5}
\newcommand{\yh}{\hat{\gamma}}
\newcommand{\yt}{\tilde{\gamma}}
\newcommand{\yb}{\bar{\gamma}}
\newcommand{\ellb}{\bar{\ell}}
\newcommand{\g}{g}
\newcommand{\gh}{\hat{g}}
\newcommand{\gt}{\tilde{g}}
\newcommand{\qbar}{\bar{q}}

\newcommand{\al}{\alpha}
\newcommand{\be}{\beta}
\newcommand{\del}{\delta}
\newcommand{\eps}{\epsilon}
\newcommand{\ve}{\varepsilon}
\newcommand{\ib}{{\bar\imath}}
\newcommand{\jb}{{\bar\jmath}}

\newcommand{\kt}{\tilde{k}}
\newcommand{\Pp}{{\cal P}_+}
\newcommand{\Pm}{{\cal P}_-}
\newcommand{\Ppm}{{\cal P}_\pm}
\newcommand{\Pmp}{{\cal P}_\mp}

\newcommand{\Nf}{N_f}
\newcommand{\Nc}{N_c}
\newcommand{\cg}{c_\Gamma}

\newcommand{\ycut}{y_{\rm cut}}


\begin{titlepage}
\vspace*{-2cm}
\begin{flushright}
SLAC--PUB--7309\\
hep-ph/9609460\\
September, 1996\\
\end{flushright}
\vskip 2.5cm
\begin{center}
{\Large\bf
Electron-Positron Annihilation into Four Jets
} \\
\vskip 0.2cm
{\Large\bf
at Next-to-Leading Order in $\alpha_s$
}\footnote{
Research supported by the Department of
Energy under grant DE-AC03-76SF00515, 
and by the Swiss National Science Foundation}\\
\vskip 1cm
{\large Adrian Signer and Lance Dixon} \\
\vskip 0.3cm
{\it 
Stanford Linear Accelerator Center \\
Stanford University \\
Stanford, CA 94309
} \\
\vskip 2cm
\end{center}

\begin{abstract}
\noindent
We calculate the rate for $e^+e^-$ annihilation into four jets
at next-to-leading order in perturbative QCD, 
but omitting terms that are suppressed by one or more powers of 
$1/\Nc^2$, where $\Nc$ is the number of colors.  
The $\cO(\alpha_s^3)$ corrections depend strongly on 
the jet resolution parameter $\ycut$ and on the
clustering and recombination schemes, and they substantially 
improve the agreement between theory and data.
\end{abstract}

\vskip 2cm
\begin{center}
{\sl Submitted to Physical Review Letters}
\end{center}

\end{titlepage}
\setcounter{footnote}{0}

\bigskip


Jets have proven to be an extremely useful way to describe 
the production of hadrons in $e^+e^-$ annihilation and 
in hadronic collisions containing large transverse momenta.
Since jets can be given an infrared-safe definition~\cite{JetDef}, 
their properties can be calculated in perturbative quantum
chromodynamics (QCD), order-by-order in the strong coupling $\alpha_s$.  
Electron-positron annihilation provides the cleanest experimental 
situation for studying jet properties, and large data samples from
the $Z^0$ pole are available.  
On the theoretical side, there are leading order predictions for 
production of up to five 
jets~\cite{ThreeJetsBorn,FourJetsBorn,ERT,FSKS,FiveJetsBorn},
but to improve the precision next-to-leading-order QCD corrections 
are required.
The $\cO(\alpha_s^2)$ matrix elements for three-jet production 
and other infrared-safe quantities have been known for some 
time~\cite{ERT,FSKS}, and numerical programs implementing
these corrections~\cite{ThreeJetsPrograms}
have been widely used to extract a precise value of $\alpha_s$ from
hadronic event shapes~\cite{EventShapeAlphas}.  

Four-jet final states provide certain tests of QCD to which three-jet 
states are insensitive~\cite{FourJetTests}.  
For example, the non-abelian three-gluon
vertex appears at leading order in four jet events;
the same is true for the production of hypothetical, light, colored 
but electrically neutral particles (such as light gluinos).
In addition, four jet events produced directly in annihilation form a
significant experimental background to the reaction $e^+e^- \to W^+W^-$
when each $W$ decays to a pair of jets, particularly when the 
center-of-mass energy is not far above the $W$-pair threshold, 
as is the case at LEP2.
In this letter, we report the next-to-leading-order ($\cO(\alpha_s^3)$) 
QCD predictions for $e^+e^-$ annihilation into four jets, 
in a large-$\Nc$ approximation to be described shortly.  
The results rely heavily on the one-loop virtual matrix elements
for four massless final state partons, $e^+e^-\to q\qbar gg$ and 
$e^+e^-\to q\qbar q'\qbar'$~\cite{Zqggq}
(for an independent calculation of $e^+e^-\to q\qbar q'\qbar'$,
see ref.~\cite{GloverMiller}),
as well as the tree-level matrix elements for five final state partons,
$e^+e^-\to q\qbar ggg$ 
and $e^+e^-\to q\qbar q'\qbar'g$~\cite{FiveJetsBorn}.
Here we give results only for the overall four-jet event rate;
but the same numerical program may be used to calculate various 
angular distributions~\cite{FourJetAngles}
which may test QCD more stringently.

Our main approximation to the full $\cO(\alpha_s^3)$ QCD results 
for massless quarks consists of omitting terms that are suppressed 
by one or more powers of $1/\Nc^2$, 
where $\Nc$ is the number of colors in a general $SU(\Nc)$
gauge theory, and $\Nc=3$ for QCD.  
Thus, extracting an overall factor of $(\Nc^2-1)$ common to all 
multi-jet predictions, we write the one-loop correction
to the four-jet cross-section as
\beq
\sigma_{4-{\rm jet}}^{1-{\rm loop}} 
= \Nc^2 (\Nc^2 - 1) \left[
 \sigma_4^{(a)} + (\Nf/\Nc)\, \sigma_4^{(b)} 
 + (\Nf/\Nc)^2 \, \sigma_4^{(c)} 
 + \cO(1/\Nc^2) + \cO(\Nf/\Nc^3) \right]\,, 
\eeq
and we calculate $\sigma_4^{(a,b,c)}$.
This is not precisely the $1/\Nc$ approximation of 
't Hooft~\cite{DoubleLine},
because we keep terms that are only suppressed by $\Nf/\Nc$,
where $\Nf$ is the number of light fermions, $\Nf=5$ at the 
$Z^0$ pole.  The ratio $\Nf/\Nc$ is not small, and the $\Nf/\Nc$
corrections {\it are} numerically important.
To assess whether the omitted $1/\Nc^2$ corrections can be expected 
to be small, we have evaluated the $\cO(\alpha_s^2)$ three-jet results 
in the same approximation, but where we also know the full result.
The size of the neglected terms varies with the precise jet definition,
but it is generally about 10\% of the full $\cO(\alpha_s^2)$ correction.
It is also known that the $1/\Nc^2$ terms contribute less than 6\% to 
the $\cO(\alpha_s^3)$ term in the expansion of the total
cross-section for $e^+e^-$ to hadrons~\cite{NNLOTotalCrossSection}. 
In future work the $1/\Nc^2$ corrections should be available.
In any case, it makes sense to break up the numerical evaluation 
in this way.  The $1/\Nc^2$ corrections are significantly more
complicated than the leading terms and therefore take longer to 
evaluate numerically, yet they are parametrically suppressed.
Hence one can save computer time without sacrificing overall accuracy
if one separately evaluates the $1/\Nc^2$ corrections, using fewer 
points in their Monte Carlo integration than one uses for the 
leading terms.  We kept all $1/\Nc^2$-suppressed terms in the 
tree-level ($\cO(\alpha_s^2)$) cross-section, except for
the four-quark terms coming from Pauli-exchange 
(the ``E'' terms of ref.~\cite{ERT}), which are numerically tiny.  

We omit two other classes of $\cO(\alpha_s^3)$ contributions:
\par\noindent
1. Contributions proportional to the axial coupling $a_f$ of the $Z^0$ 
to quarks.  Analogous terms have traditionally 
been omitted from $\cO(\alpha_s^2)$ programs, as they cancel precisely 
between up- and down-type quarks in the final state (for zero quark
mass), and their contribution to the three-jet rate is at the percent
level~\cite{AxialCoupling}.
\par\noindent
2. Contributions proportional to $(\sum_f v_f)^2$, where $v_f$ is 
the quark vector coupling.   These ``light-by-glue scattering'' 
terms do not appear at $\cO(\alpha_s^2)$ at all, have a partial 
cancellation from the sum over quark flavors, and contribute less 
than 1\% to the $\cO(\alpha_s^3)$ term in the total 
cross-section~\cite{NNLOTotalCrossSection}. 

 
\figmac{3.7}{allgraphs}{AllGraphs}{} 
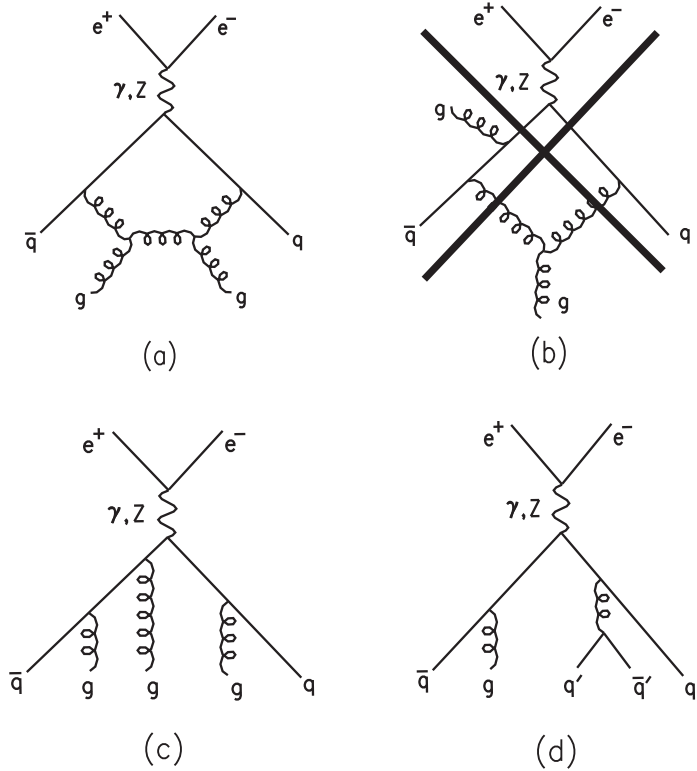
{{\small (a) Example of a leading-in-$\Nc$ one-loop diagram for 
$e^+e^-\to q\qbar gg$.  (b) A subleading one-loop diagram (omitted).
(c) Sample tree diagram for $e^+e^-\to q\qbar ggg$.
(d) Sample tree diagram for $e^+e^-\to q\qbar q'\qbar' g$.
\hfill}}


The $1/\Nc^2$ expansion is facilitated by using a color-ordered 
framework for both the one-loop and tree matrix 
elements~\cite{ColorOrder}. 
The amplitudes are decomposed into kinematical quantities ---
{\it partial amplitudes} ---
multiplying particular strings (and traces) of the fundamental 
$SU(\Nc)$ generator matrices $(T^a)_{i}^{~\jb}$.
In the expression for the cross-section, i.e. the squared amplitude, 
summed over all color indices,
it is easy to identify the partial amplitudes that accompany the 
highest powers of $\Nc$.  They are given by the sum of 
{\it color-ordered} Feynman diagrams, where the cyclic ordering 
of the external quark and gluon legs is fixed, and where the 
gluons are on the opposite side of the Feynman diagram from the
$e^+e^-$ pair.
For example, \fig{AllGraphs}a shows a diagram contributing to 
a leading-color partial one-loop amplitude for $e^+e^-\to q\qbar gg$,
while \fig{AllGraphs}b (crossed out) shows a diagram that only contributes
to subleading-color partial amplitudes.
\fig{AllGraphs}c and \fig{AllGraphs}d are sample diagrams 
for the five-parton tree amplitudes. 

As usual, the real and virtual contributions to the 
cross-section are separately divergent; only the sum 
of the two yields a meaningful finite result.  
In dimensional regularization with $D=4-2\epsilon$, the singularities 
of the virtual part manifest themselves as poles in $\epsilon$
in the one-loop amplitudes, whereas the real singularities are 
obtained upon phase-space integration of the squared tree amplitudes.
In order to combine these two contributions,
we use a general version of the subtraction 
method~\cite{ERT}. 
In fact, the program used in this letter is a straightforward 
implementation of the method developed in ref.~\cite{FKS}. 
We refer the reader to this article for more details. 
Here we only mention that no approximation of the matrix elements
has to be made, and that we checked the independence of the results 
on the arbitrary parameters $\delta$ and $\xi_{cut}$ which have to 
be introduced in intermediate steps. 


We now present the results for the four-jet fraction 
$R_4 \equiv \sigma_{4-{\rm jet}}/\sigma_{{\rm tot}}$
at next-to-leading order in $\alpha_s$.  
We consider four different jet algorithms, 
the JADE~\cite{JADE}, E0, Durham~\cite{Durham} 
and Geneva~\cite{schemes} schemes.\footnote{
Lacking a scheme to name after our fair city, 
we have dubbed our numerical program 
Matrix Elements for Next-to-Leading Order PARton Calculations 
(MENLO\_PARC).}  
All four are iterative clustering algorithms:
they begin with a set of final-state particles (partons in the
QCD calculation) and cluster the pair $\{i,j\}$ with the 
smallest value of a dimensionless measure $y_{ij}$ into a single
``proto-jet''.  The procedure is repeated until all the $y_{ij}$ 
exceed the value of the jet resolution parameter $\ycut$, at which
point the proto-jets are declared to be jets.
The schemes differ in the measure $y_{ij}$ used and/or in the rule 
for recombining two clustered momenta.
The same value of $\ycut$ in different schemes may sample different 
momentum scales.  
The definitions of the various schemes are collected in 
ref.~\cite{schemes}. 
The $\ycut$ dependence of the results is shown in \fig{ycutdep}.
In each plot, the solid (dashed) line represents the one-loop 
(tree-level) prediction. 
The renormalization scale $\mu$ has been chosen to be the
center-of-mass energy $\sqrt{s}$, 
and we have set $\Nf=5$ and $\al_s = 0.118$~\cite{AverageAlphas}. 
The statistical error of the Monte Carlo integration is of the 
order of 3\%. 
Of course, this error can be reduced further through higher
statistics, but as long as the subleading-in-$\Nc$ terms are not 
included, there is little point in doing so.
These curves are compared to preliminary SLD data 
points~\cite{SLDdata} which have been corrected for detector 
effects and hadronization, 
and to available LEP1 data~\cite{LEPdata}, which have not
been corrected for hadronization. 

 
\figmac{5.0}{R4}{ycutdep}{}  
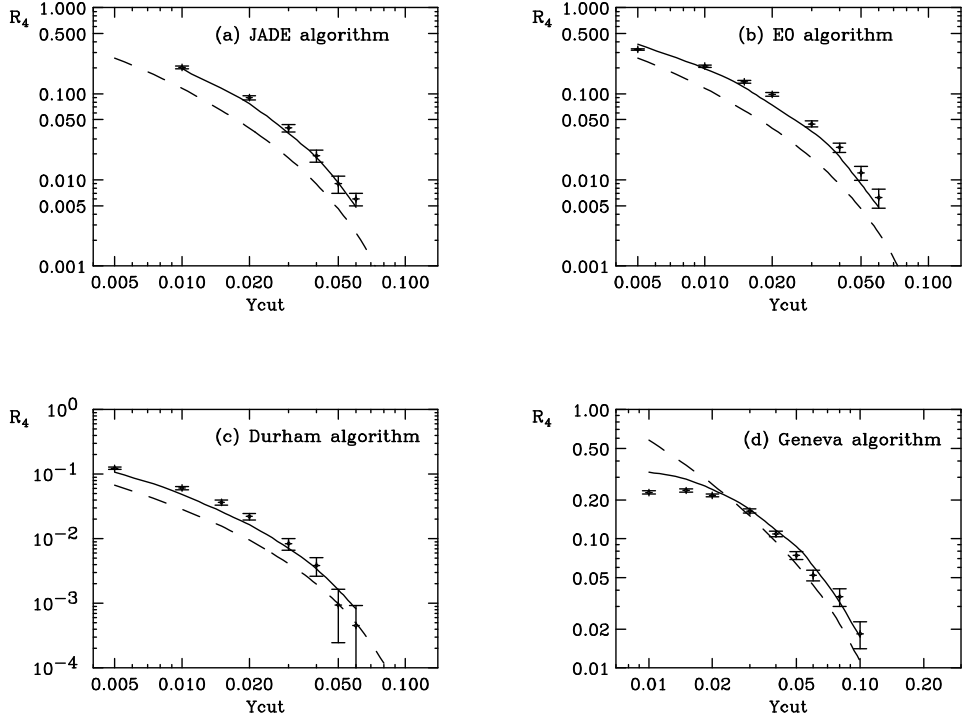
{{\small The four-jet fraction $R_4$ in $e^+e^-$ annihilation, 
as a function of $\ycut$.  
Solid (dashed) lines represent the one-loop (tree-level) 
predictions in the (a) JADE, (b) E0, (c) Durham, and (d) Geneva
algorithms, for $\mu=\sqrt{s}$ and $\al_s = 0.118$.
The data points in (a) are from DELPHI~\cite{LEPdata}
(uncorrected for hadronization),
while (b), (c) and (d) contain preliminary SLD 
data~\cite{SLDdata} (corrected for hadronization).
\hfill}}

 
\figmac{5.0}{mu}{mudep}{}  
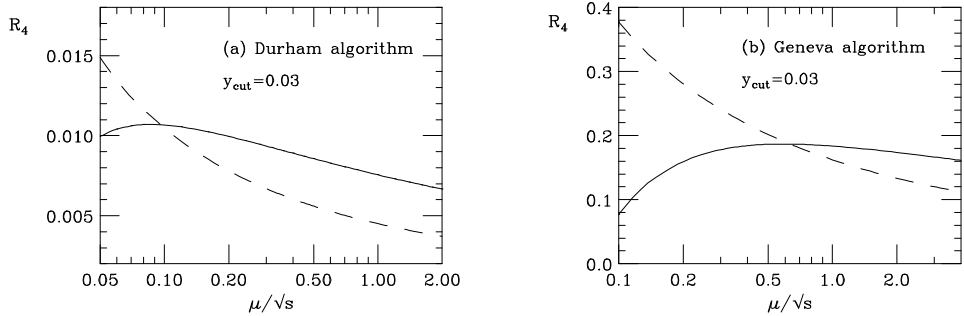
{{\small Solid (dashed) lines show the dependence of $R_4$
on the renormalization scale $\mu$ 
for the one-loop (tree-level) predictions in the 
(a) Durham, and (b) Geneva algorithms, for $\al_s = 0.118$ 
and $\ycut = 0.03$.
\hfill}}


The next-to-leading order results in the JADE algorithm in 
\fig{ycutdep}a actually agree ``too well'' with the data, 
given that the one-loop corrections are of order 100\%
(and that the JADE data are uncorrected for hadronization).
The results in the E0 and Durham schemes (\fig{ycutdep}b,c)
are more what one might expect
from such large corrections: agreement to within 20 or 30\%.
The Geneva scheme (\fig{ycutdep}d) behaves quite differently.
It is the only scheme we considered where the leading-order
results give a reasonable description of the data for large values of
$\ycut$, although the shape of the prediction is not quite correct,
especially at small $\ycut$.  Here the inclusion of the
one-loop correction leads to quite good agreement between data and 
theory for $\ycut > 0.02$.  However, for $\ycut < 0.02$, 
the one-loop virtual corrections become very large and negative
(they are dominated by the $\Nf/\Nc$ terms in this region), but 
not enough to match the strong suppression and turnover seen in 
the four-jet data.
In any scheme, as $\ycut$ decreases the four-jet fraction rises
quickly, but eventually it has to turn over; this phenomenon just 
happens at a larger value of $\ycut$ in the Geneva scheme than in 
the other schemes.
We have also compared the one-loop prediction at $\sqrt{s}=35$~GeV 
to data from PETRA~\cite{PETRAdata}.  
The agreement is again improved with respect to the leading-order 
result, but it is not quite as good as at the $Z^0$ pole,
perhaps because hadronization effects are more important 
at lower energy.

The rapid fall-off of the four-jet fraction at large $\ycut$
means that there is little data available (at present) with 
which to compare our predictions for $\ycut > 0.07$.  
On the other hand, for small $\ycut$ 
the QCD expansion parameter is really $\al_s \ln^2\ycut$, 
and it would be advantageous to resum these large logarithms.  
This is possible at leading and next-to-leading logarithmic order 
in the Durham scheme~\cite{Durham}.  
To further improve the Durham-scheme prediction, our fixed-order 
results could be matched~\cite{Matching} to the resummed results. 

Observable quantities calculated in QCD should be independent of the
arbitrary renormalization scale $\mu$.  However, the perturbative 
expansion is invariably truncated at a finite order, leading
to a residual dependence of the result on $\mu$.
The tree-level $\mu$ dependence is much stronger for the four-jet rate 
than for the three-jet rate, because the former is proportional to
$\al_s^2$ instead of $\al_s$. 
As expected, this strong renormalization-scale dependence is reduced
by the inclusion of the next-to-leading order contribution.   
\fig{mudep} plots the $\mu$-dependence of $R_4$ at tree-level
and at one-loop for the Durham and Geneva schemes, at $\ycut = 0.03$. 
We should mention that in order to get a consistent picture, 
for these plots alone we omitted the corresponding $1/\Nc^2$ corrections 
from the tree-level term as well.  On the other hand, their inclusion 
only slightly affects the $\mu$-dependence. 


In this letter we have presented first results on the production
of four jets in electron-positron annihilation at 
next-to-leading order in $\al_s$.  
These results were obtained with a numerical program which implements 
the subtraction method for combining real and virtual singularities
as described in~ref.~\cite{FKS}. 
The key ingredients for the four-jet calculation are the one-loop
virtual matrix elements for $e^+e^-\to q\qbar gg$ and 
$e^+e^-\to q\qbar q'\qbar'$~\cite{Zqggq}, as well as the tree-level matrix
elements for $e^+e^-\to q\qbar ggg$  and 
$e^+e^-\to q\qbar q'\qbar'g$~\cite{FiveJetsBorn}.
We computed the four-jet rates for the JADE, E0, Durham and Geneva 
schemes, neglecting terms that are suppressed by $1/\Nc^2$. 
Generally, the corrections are large and improve the agreement between 
theory and experiment considerably. 
The same program can be used for the computation of an {\it arbitrary}
four-jet distribution at next-to-leading order. 
In future work we shall report on the computation of various 
angular distributions, including also in the program
the subleading-in-color terms. 
These calculations may help in testing the triple gluon vertex 
and in the more general search for new physics in the strong 
interaction sector.


\begin{flushleft}
{\bf\large Acknowledgement} \\
\end{flushleft}

We thank Zvi Bern, Phil Burrows, David Kosower and Zoltan Kunszt
for valuable conversations.

\newpage


\def\np#1#2#3  {{ Nucl. Phys. }{#1}:{#2} (19#3)}
\def\nc#1#2#3  {{ Nuovo. Cim. }{#1}:{#2} (19#3)}
\def\zp#1#2#3  {{ Z. Phys. }{#1}:{#2} (19#3)}
\def\pl#1#2#3  {{ Phys. Lett. }{#1}:{#2} (19#3)}
\def\pr#1#2#3  {{ Phys. Rev. }{#1}:{#2} (19#3)}
\def\prl#1#2#3  {{ Phys. Rev. Lett. }{#1}:{#2} (19#3)}
\def\prep#1#2#3  {{ Phys. Rep. }{#1}:{#2} (19#3)}

\end{document}